\documentclass[twocolumn,showpacs,preprintnumbers,
prb,superscriptaddress]{revtex4-1}%
\usepackage{amssymb}
\usepackage{mathtools}
\usepackage{amsmath}
\usepackage{graphicx}
\usepackage{float}
\usepackage{dcolumn}
\usepackage{bm}
\usepackage[export]{adjustbox}
\usepackage{amsfonts}%
\usepackage{lipsum}% http://ctan.org/pkg/lipsum
\usepackage[usenames, dvipsnames]{color}
\providecommand{\U}[1]{\protect\rule{.1in}{.1in}}
\definecolor{greenish}{rgb}{0.20, 0.65, 0.25}

\newcommand*{\citen}[1]{%
  \begingroup
    \romannumeral-`\x % remove space at the beginning of \setcitestyle
    \setcitestyle{numbers}%
    \cite{#1}%
  \endgroup   
}

\begin{document}

%%\setcounter{MaxMatrixCols}{10}
%%\input{tcilatex}
%%\documentclass[onecolumn,showpacs,preprintnumbers,
%%prb,superscriptaddress]{revtex4}%

%%\documentclass[twocolumn,prb,aps]{revtex4}%
%\documentclass[article]{revtex4}%
%\usepackage{amssymb}
%\usepackage{amsmath}
%\usepackage{graphicx}
%\usepackage{dcolumn}
%\usepackage{bm}
%\usepackage{amsfonts}%
%\setcounter{MaxMatrixCols}{30}
%%TCIDATA{OutputFilter=latex2.dll}
%%TCIDATA{Version=5.50.0.2952}
%%TCIDATA{LastRevised=Saturday, August 05, 2017 18:57:12}
%%TCIDATA{<META NAME="GraphicsSave" CONTENT="32">}
%%TCIDATA{<META NAME="SaveForMode" CONTENT="1">}
%%TCIDATA{BibliographyScheme=Manual}
%%BeginMSIPreambleData
%\providecommand{\U}[1]{\protect\rule{.1in}{.1in}}
%%EndMSIPreambleData
%\providecommand{\func}[1]{\textrm{#1}}
%\begin{document}
\title{Correlation Effects in Orbital Magnetism}
\author{T. J. Sj\"{o}strand}
\affiliation{Department of Physics, Division of Mathematical Physics, Lund
 University,
Professorsgatan 1, 223 62 Lund, Sweden}
\author{K. Karlsson}
\affiliation{Department of Engineering Sciences, University of Sk\"{o}vde, SE-541 28
Sk\"{o}vde, Sweden}
\author{F. Aryasetiawan}
\affiliation{Department of Physics, Division of Mathematical Physics, Lund University,
Professorsgatan 1, 223 62 Lund, Sweden}
\date{\today }

\begin{abstract}
Orbital magnetization is known empirically to play an important role in several magnetic phenomena, such as permanent magnetism and ferromagnetic superconductivity. Within the recently developed ''modern theory of orbital magnetization'', theoretical insight has been gained into the nature of this often neglected contribution to magnetism, but is based on an underlying mean-field approximation. From this theory, a few treatments have emerged which also take into account correlations beyond the mean-field approximation.  Here, we apply the scheme developed in a previous work [Phys. Rev. B \textbf{93}, 161104(R) (2016)] to the Haldane-Hubbard model to investigate the effect of charge fluctuations on the orbital magnetization within the $GW$ approximation. Qualitatively, we are led to distinguish between two quite different situations: (i) When the lattice potential is larger than the nearest neighbor hopping, the correlations are found to boost the orbital magnetization. (ii) If the nearest neighbor hopping is instead larger than the lattice potential, the correlations reduce the magnetization. 
%We identify a strong effect from the staggered ionic potential on both the mean-field potential and on the correlations. For a staggered potential larger than the nearest neighbor hopping, the $GW$ correlations are found to increase the magnetization for physical values of on-site $U$. 
%%%%The role of charge correlations is {\krister{mainly studied} in two cases, where (i) a large staggered potential between sublattice A and B is present, and, (ii) where it vanishes. In case (i), the $GW$ correlations increase the magnetization for on-site $U$ of the order of the nearest-neighbor hopping, and in case (ii), the correlations instead decrease the magnetization.  
\end{abstract}

\maketitle

\section{Introduction}
Magnetism is usually associated with the spin of the electrons since the magnetic moment arising from the orbital motion of the electrons is normally weak or even completely quenched crystals with time-reversal symmetry. However, the role of the orbital motion in determining magnetic properties is not to be underestimated. Magnetic anisotropy, which is important in permanent magnets, has its roots in the orbital magnetic moment.\cite{hjort1997,kan2016,schoen2017} Magnetic susceptibility,\cite{wang2007} magnetoelectric response\cite{mala2010, essin2009, essin2010} and
spin-dependent transport\cite{xiao2006, murakami2006, xiao2007, wang2007} can also be strongly influenced by it. Recent experiments on ferromagnetic superconductors show the possibility of the orbital moment dominating over the spin counterpart.\cite{wilhelm2017} This is particularly notable as the two appear with opposite signs and tend to cancel each other. 

The orbital magnetic moment in a crystal has two contributions: one arising from the local atomic orbital moments and, less obvious, a contribution from the itinerant edge states on the surface of the crystal, which in the thermodynamic limit can be expressed solely as bulk quantities.\cite{thonhauser2005,thonhauser2006,ceresoli2006,xiao2005} Computing the orbital magnetization in infinite crystals has long been recognized to be nontrivial due to the position operator being ill-defined in the extended Bloch basis.\cite{resta1998, resta2005} Not long ago, the so called {\it modern} theory of orbital magnetization was established,\cite{shi2007, thonhauser2011, resta2010, xiao2010} where the orbital magnetization was expressed as a true bulk quantity evaluated from the extended Bloch states. These pioneering works have opened up a way for computing the orbital magnetic moment from realistic band structure calculations,\cite{ceresoli2010,lopez12} usually obtained from Kohn-Sham density functional theory.
%With the exception of iron, with delocalized $3d$ electrons, the itinerant contribution is rather small. Perhaps
%this is not entirely surprising since the electrons associated with the
%orbital magnetization are rather localized. Indeed, the $3d$ electrons in iron
%are relatively less localized than in nickel and cobalt, resulting in larger
%itinerant contribution. 

An interesting application of the modern theory is to materials where the electrons responsible for the orbital magnetization are semi-itinerant. A well-known example is iron, which, contrary to nickel and cobalt, has a large itinerant contribution coming from the $3d$ electrons. Recently, Hanke \textit{et al}.\cite{hanke2016} applied the theory to a variety of structural inhomogeneous systems, including topological ferromagnets. Several prototypical insulating perovskite transition metal oxides have also beed studied.\cite{nikolaev2014} Magnetic thin films,\cite{miron2011} non-collinear spin systems and frustrated spin lattices are additional candidates with spintronics applications.\cite{taguchi2001, heinze2011}

The modern mean-field theory is not expected to yield physically meaningful results for moderately to strongly correlated materials. Even in the ferromagnetic transition metals, the orbital moment is known to be underestimated when performing density functional calculations using the gradient corrected PBE functional.\cite{ceresoli2010} Systematic improvement of the calculation of the orbital magnetization in correlated systems is particularly crucial when the desired physical quantity depends on the balance between the orbital and spin magnetization. An important example is the compensation temperature of zero-magnetization ferromagnets, at which the spin and orbital magnetization exactly cancel out.

Extensions have been formulated within current and spin density functional theory,\cite{shi2007} but also in terms of the full Green's function and vertex function.\cite{kotliar2014} Recently, we have derived a formula from {\it first principles} for the orbital magnetic moment of interacting electrons.\cite{aryasetiawan2016,aryasetiawan2017} The formula factorizes into two parts, one that contains the information about the one-particle band structure and another that contains the effect of exchange and correlations beyond the local density approximation, carried by the Green's function. The self-energy, which determines the Green's function via Dyson's equation, can be calculated using the \emph{GW} approximation \cite{hedin1965} for weakly to moderately correlated materials and the LDA+DMFT\cite{georges1996} or \emph{GW}+DMFT\cite{biermann2003} approximations for strongly correlated materials. 

%On the other hand, a system respecting time-reversal symmetry will in addition to spin-orbit interactions and inversion symmetry breaking,
%give rise to $Z_2$ topological insulators\cite{kane2005} or the so called quantum spin Hall effect (QSHE). These insulators are characterized by the number of Kramer pairs at each edge of the system, where the topological non-trival (trivial) insulator has an odd (even) number of pairs, respectively. A topological phase-transition is accomplished by a gap-closing similar to the Chern insulator. A strong candidate for observing the QSHE is graphene, although the spin-orbit interaction seems to be too week. Experimentally, the QSHE was first observed in 2D CdTe/HgTe/CdTe quantum wells\cite{konig2007}, but has been found more recently also in 3D systems.\cite{zhang2009,xia2009,chen2009}  
%As previously discussed, the modern theory for calculating the orbital magnetic moment is based on non-interacting electrons and it is indeed a %challenge to incorporate electron-electron interactions from \textit{first principles}. 
\begin{figure}[h!]
\includegraphics[width=1\linewidth]{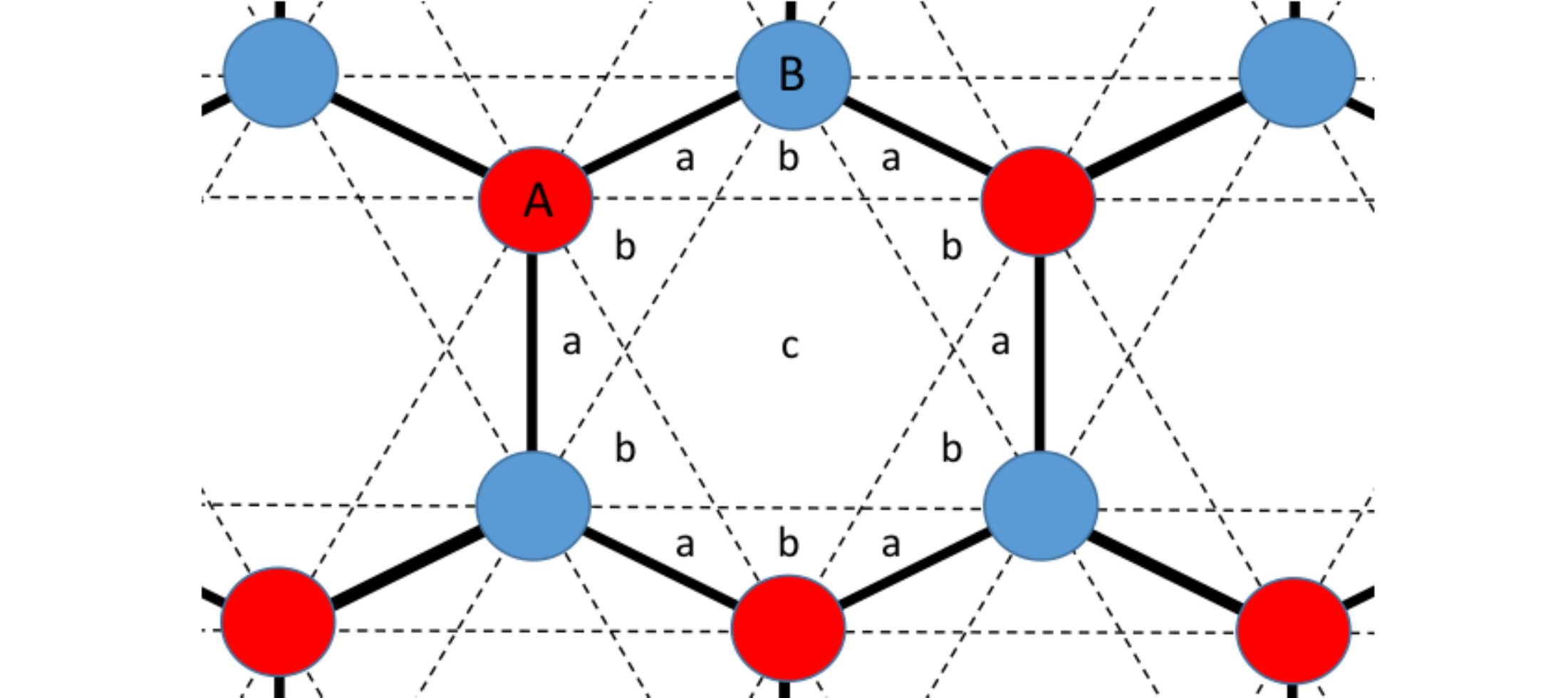}
\caption{The honeycomb lattice with two different (A and B) atoms per unit cell. The fluxes in regions $a$, $b$ and $c$ obey $\varphi_{a} = -\varphi_{b}$ and $\varphi_{c} = 0$. This implies that the total flux in the unit cell is zero. The flux $\varphi$ in Eq. \eqref{H0} is defined as $\varphi= 2\varphi_a + \varphi_b$. The figure is from Ref. \protect\citen{aryasetiawan2017}.} \label{honey}
\end{figure}

Here, we apply our previously derived Green's function formula for the orbital magnetization to the 2D Haldane-Hubbard model within the $GW$ approximation, as a first step to understand the effect of correlations on the magnetization in real systems. In Sec. II we briefly describe the essentials of the Haldane-Hubbard model. In Sec. III we revisit the - until now - untested Green's function formula and summarize the approximations and computational details of this work. We present the results within the $GW$ and Hartree-Fock approximations in Sec. IV for different values of the parameters of the model. The main findings are discussed in Sec. V.   

\section{The 2D Haldane-Hubbard Model}
The famous 2D Haldane model \cite{haldane1988} has been used for decades in studies of topologically non-trivial systems, such as 
Chern insulators and $Z_2$ topological insulators.\cite{kane2005} The Haldane model mimics the quantum Hall effect\cite{klitzing1980} with a time-reversal breaking internal microscopic field which averages to zero in the unit cell. Recently, the Haldane model was experimentally realized by using ultra-cold atoms.\cite{jotzu2014}

In the non-interacting Haldane model the Hamiltonian is expressed as 
\begin{align}
\hat{H}_0  =& -t_1 \sum_{i\sigma} \big[ a_{i\sigma}^\dag b_{i+{\boldsymbol \delta}_1\sigma } \!+\! a_{i\sigma}^\dag b_{i+{\boldsymbol \delta}_2 \sigma} \!+\! a_{i\sigma}^\dag b_{i+{\boldsymbol \delta}_3 \sigma} \!+\! \text{h.c.}\big]  \nonumber \\
& -t_2 \big( e^{i\varphi} \sum_{i\sigma} \big[ a_{i\sigma}^\dag a_{i+{\boldsymbol \nu}_1 \sigma} \!+\! a_{i+{\boldsymbol \nu}_1 \sigma}^\dag a_{i-{\boldsymbol \nu}_2 \sigma} \!+\! a_{i-{\boldsymbol \nu}_2 \sigma}^\dag a_{i\sigma} \big] \nonumber \\
&+ \text{h.c.} \!+\! \big[ \varphi \!\to\! -\varphi,a \!\to\! b \big] \big) \!+\! \Delta_\text{AB}\sum_{i \sigma} \big[ a^{\dagger}_{i \sigma }  a_{i\sigma}
\!-\!  b^{\dagger}_{i\sigma}  b_{i\sigma} \big] , \label{H0}
\end{align}
%\begin{align}
%  \hat{H}_0 =& -t_1 \! \sum_{\langle ij \rangle \sigma } \! \big[ a^{\dagger}_{i\sigma}  b_{j\sigma}+
%  b_{j\sigma}^\dag a_{i\sigma}
%  \big] \nonumber \\
%  & - t_2 \!\sum_{\langle \langle ij \rangle \rangle \sigma } \big[\big( e^{i\varphi}a^{\dagger}_{i\sigma} a_{j\sigma} \! + \!  e^{-i\varphi}b^{\dagger}_{i\sigma}  b_{j\sigma} \! \big) + \text{h.c.}\big]  \nonumber \\ &
%+ \Delta_\text{AB}\sum_{i \sigma} \big[ a^{\dagger}_{i \sigma }  a_{i\sigma}
%-  b^{\dagger}_{i\sigma}  b_{i\sigma} \big] , \label{H0}
%\end{align}
The annihilation operators $a_{i\sigma}$ belong to sublattice A and $b_{i\sigma}$ to sublattice B. We restrict our study to having two electrons per unit cell, which results in one fully occupied and one empty band. $t_1$ is the real nearest-neighbor hopping between sublattices A and B of the honeycomb lattice. With lattice constant $a=1$, the nearest neighbor vectors are ${\boldsymbol \delta}_1 = (0,1)$, ${\boldsymbol \delta}_2 = (-\frac{\sqrt{3}}{2},\frac{1}{2})$ and ${\boldsymbol \delta}_3 =  (\frac{\sqrt{3}}{2},\frac{1}{2}) $, see Fig. \ref{honey}. The next-nearest neighbor vectors that explicitly enter the Hamiltonian are ${\boldsymbol \nu}_1 =(\sqrt{3},0)$ and ${\boldsymbol \nu}_2 = (\frac{\sqrt{3}}{2}, - \frac{3}{2})$. $t_2$ and $\varphi$ denote the amplitude and complex phase of the next-nearest neighbor hopping. $\varphi$ mimics a microscopic magnetic field which breaks time-reversal symmetry if it differs from a multiple of $\pi$. The magnetic flux, by construction, averages to zero in the unit cell. $\Delta_\text{AB}$ models a staggered potential between the two sublattices and breaks the inversion symmetry and gives mass to the Dirac fermions. Pure graphene has $\Delta_\text{AB}=0$ whereas epitaxial graphene has a non-vanishing $\Delta_\text{AB}$ (see e.g. Ref. \citen{ci2010}). The possibility of tuning the band gap in graphene by doping makes way for a wider range of technological applications, e.g. logic gates. The effect of this doping is beautifully captured by the Haldane model, but it is also ideal for describing several insulators, with a larger value of $\Delta_\text{AB}$, such as two-dimensional hexagonal boron nitride.

The Hall conductance $\sigma_{xy}$ is given by an integer $C$, called the Chern number, times $e^2/h$. $C$ is an intrinsic property of the band structure, given by
\begin{align}
C= \frac{i}{2\pi}\! \sum_n \int_{BZ} \! \frac{d{\bf q}}{(2\pi)^2} \! \bigg[\!\bigg\langle \frac{\partial u_{n{\bf q}}}{\partial q_x }\bigg|\frac{\partial u_{n{\bf q}}}{\partial q_y } \bigg\rangle \!- \! \bigg\langle \frac{\partial u_{n{\bf q}}}{\partial q_y }\bigg|\frac{\partial u_{n{\bf q}}}{\partial q_x } \bigg\rangle \! \bigg]  ,   
\end{align}
for periodic 2D systems, where $|u_{n{\bf q}} \rangle= \text{e}^{-i{\bf q}\cdot {\bf r}} |\psi_{n{\bf q}} \rangle$ are the Bloch-periodic functions. Note that $C$ is independent of the gauge chosen for $|u_{n{\bf q}} \rangle$. This formula is also valid when including correlations, but $|u_{n{\bf q}}\rangle$ will then contain additional information, as discussed in the next section. The phase transitions where $C$ changes are always accompanied by the closing and reopening of the band gap. 
%\tor{Vertical gap, be more specific!} 

The total Hamiltonian, $\hat{H}=\hat{H}_0 +\hat{V}_{e-e}$, also contains the electron-electron interaction term 
\begin{align}
\hat{V}_{e-e} & = \frac{1}{2} \sum_{ijkl, \sigma \sigma'} U_{ijkl} c_{i\sigma}^\dag c_{j\sigma'}^\dag c_{l\sigma'} c_{k\sigma}, \label{vee}
\end{align}
where $c_i=a_i$ or $c_i=b_i$ if $i$ corresponds to sublattice A or B, respectively, and 
\begin{align}
    U_{ijkl} & = \int d {\bf r} d {\bf r}' \frac{\phi_i^*({\bf r}) \phi_j^*({\bf r}')\phi_l({\bf r}')\phi_k({\bf r})}{|{\bf r}-{\bf r}'|} .
\end{align}
In this work, the orbitals $\phi_i$ are assumed to be localized and only the short-ranged direct Coulomb integrals $U=U_{iiii}$ and $U'=U_{ijij}$ are kept, where $i$ and $j$ are nearest neighbor sites. $U$ is the on-site interaction of the Hubbard model and $U'$ is the nearest-neighbor interaction. We intend to go beyond this approximation in a future study on real materials by including more Coulomb integrals, in particular exchange integrals $U_{ijji}$, all calculated using the constrained random-phase approximation.\cite{crpa} Indeed, this approximation has already been applied to graphene, see Ref. \citen{graphene}. 
 \section{Correlations and Orbital Magnetization}\label{corr}
\subsection{Exact Expressions}
The expression for the orbital magnetization, $M_\text{orb}$, derived in Ref. \citen{aryasetiawan2016}, is readily generalized to electrons with spin when spin-orbit interaction is neglected. For spin-independent systems, like that described by Eq. \eqref{H0}, the contributions to $M_\text{orb}$ from spin $\uparrow$ and $\downarrow$ are the same. We adopt the notation that the orbital magnetization in one spin channel is $M_\text{orb}$. 

The orbital magnetization is perpendicular to the 2D plane of the Haldane model and is a sum of a local and an itinerant contribution: 
    \begin{widetext}
    \begin{align}
 M_{z}^{L}  & =   \sum_{jk,nn'}  \epsilon_{zjk}  \int_{BZ} 
\frac{d{\bf q}}{(2\pi)^{2}}  \bigg[ G^+_{n^{\prime}n}(\mathbf{q})  \left\langle
\frac{\partial u_{n\mathbf{q}}}{\partial q_{j}}\right\vert H(\mathbf{q})  - \mu\left\vert \frac{\partial u_{n^{\prime}\mathbf{q}}}{\partial
q_{k}}\right\rangle + \frac{\partial G^+_{n^{\prime}n}(\mathbf{q})}{\partial q_{k}} \left\langle
\frac{\partial u_{n\mathbf{q}}}{\partial q_{j}}\bigg|u_{n^{\prime}\mathbf{q}%
}\right\rangle (E_{n^{\prime}\mathbf{q}} - \mu) \bigg], \label{mlq} \\
 M^{I}_z  & =\sum_{jk,nn'}\epsilon_{zjk}\int_{BZ}%
\frac{d{\bf q}}{(2\pi)^{2}}G^+_{n^{\prime}n}(\mathbf{q})  \left[  \left\langle \frac{\partial u_{n\mathbf{q}}}{\partial q_{j}%
}\bigg|\frac{\partial u_{n^{\prime}\mathbf{q}}}{\partial q_{k}}\right\rangle
(E_{n^{\prime}\mathbf{q}}-\mu)  +\frac{\partial E_{n\mathbf{q}}^{0}%
}{\partial q_{j}}\left\langle \frac{\partial u_{n\mathbf{q}}}{\partial q_{k}%
}\bigg|u_{n^{\prime}\mathbf{q}}\right\rangle \right]. \label{miq}
\end{align}
\end{widetext}
The orbital magnetic
moment is obtained by multiplying with the factor $-e/2c$, in
Gaussian units. The sums are over \textsl{all} bands $n$ and $n'$. $\mu$ denotes the chemical potential and $\epsilon_{ijk}$ the Levi-Civita symbol, where $ijk$ are cartesian indices and, in particular, $z$ is perpendicular to the 2D plane. The spin index has been left out for simplicity. More precisely, $H({\bf q}) = e^{-i\mathbf{q\cdot r}}H e^{i\mathbf{q\cdot r}}$ with $H({\bf q}) | u_{n\mathbf{q}} \rangle = E_{n\mathbf{q}} | u_{n\mathbf{q}} \rangle $ where the one-particle Hamiltonian $H$ is defined from 
\begin{align}
\hat{H}_\text{MF} &= \sum_\sigma \int d{\bf r} \psi_\sigma^\dag ({\bf r}) H({\bf r}) \psi_\sigma({\bf r}) .
\end{align}
Here, $\hat{H}_\text{MF}$ is the self-consistent mean field of $\hat{H}_0 + \hat{V}_{e-e}$. 

The central quantities for the calculation of $M_\text{orb}$ are the band-index matrix elements of the one-particle Green's function,
\begin{align}
G^+_{nn'}({\bf q}) &  = \lim_{\eta \to 0^+} \int \frac{d \omega}{2\pi}  G_{nn'}({\bf q};\omega)\text{e}^{i \omega \eta}
, \label{intfreq} \\
\big[G^{-1}\big]_{nn'}({\bf q}; \omega) & =\big[ g^{-1} \big]_{nn}({\bf q};\omega)  \delta_{nn'} - \Sigma_{nn'}({\bf q};\omega)  .  \label{Dyson}
\end{align}
The Hartree-potential is contained in the mean-field Green's function, $g$, whereas the self-energy $\Sigma= \Sigma^x + \Sigma^c$ contains all effects from exchange and correlations. The Chern number for the correlated system is obtained by defining $|u_{n{\bf k}} \rangle$ from the diagonalization of $ -G^{-1}({\bf q};0)= H({\bf q}) + \Sigma({\bf q};0)$.\cite{vanhala2010} The model contains only one orbital per lattice site, so the exchange vanishes when the Coulomb interaction is approximated by the on-site $U$.  Since we assume localized orbitals, the exchange between different sites also vanishes, implying that $\Sigma^x=0$.

The main purpose of this paper is to study electron correlation effects in the context of orbital magnetization. 
We compare $M_\text{orb}$ within (a) the Hartree-Fock approximation and (b) the $GW$ approximation, for a range of value of the Coulomb integrals, $U$ and $U'$.
\subsection{Approximating $G$}\label{app}
(a) Hartree-Fock approximation: This amounts to neglecting correlations,  $\Sigma^c=0$. Since also $\Sigma^x=0$, this is identical to the Hartree approximation, so we can write $g({\bf q};\omega) = (\omega - H({\bf q}))^{-1}$.

(b) $GW$ approximation: 
\begin{align}
\Sigma^c_{nn'}({\bf q};\omega)  =  i \sum_{{\bf k}m} & \int  \frac{d\omega'}{2\pi}\big[ g_{mm}({\bf k};\omega+\omega') \nonumber
\\ & \times \langle \psi_{n{\bf q}} \psi_{m{\bf k}}|W^c(\omega') |\psi_{m{\bf k}} \psi_{n'{\bf q}} \rangle \big] ,
\end{align}
where
\begin{align}
 &    \langle \psi_{n{\bf q}} \psi_{m{\bf k}}|W^c(\omega) |\psi_{m{\bf k}} \psi_{n'{\bf q}} \rangle   = \nonumber \\
  &  \int d{\bf r}d{\bf r}'  \psi_{n{\bf q}}^*({\bf r}) \psi_{m{\bf k}}^*({\bf r}') W^c({\bf r}{\bf r}';\omega)  \psi_{n'{\bf q}}({\bf r}') \psi_{m{\bf k}}({\bf r}) .
\end{align}
\begin{figure}[h!]
\includegraphics[width=0.75\linewidth]{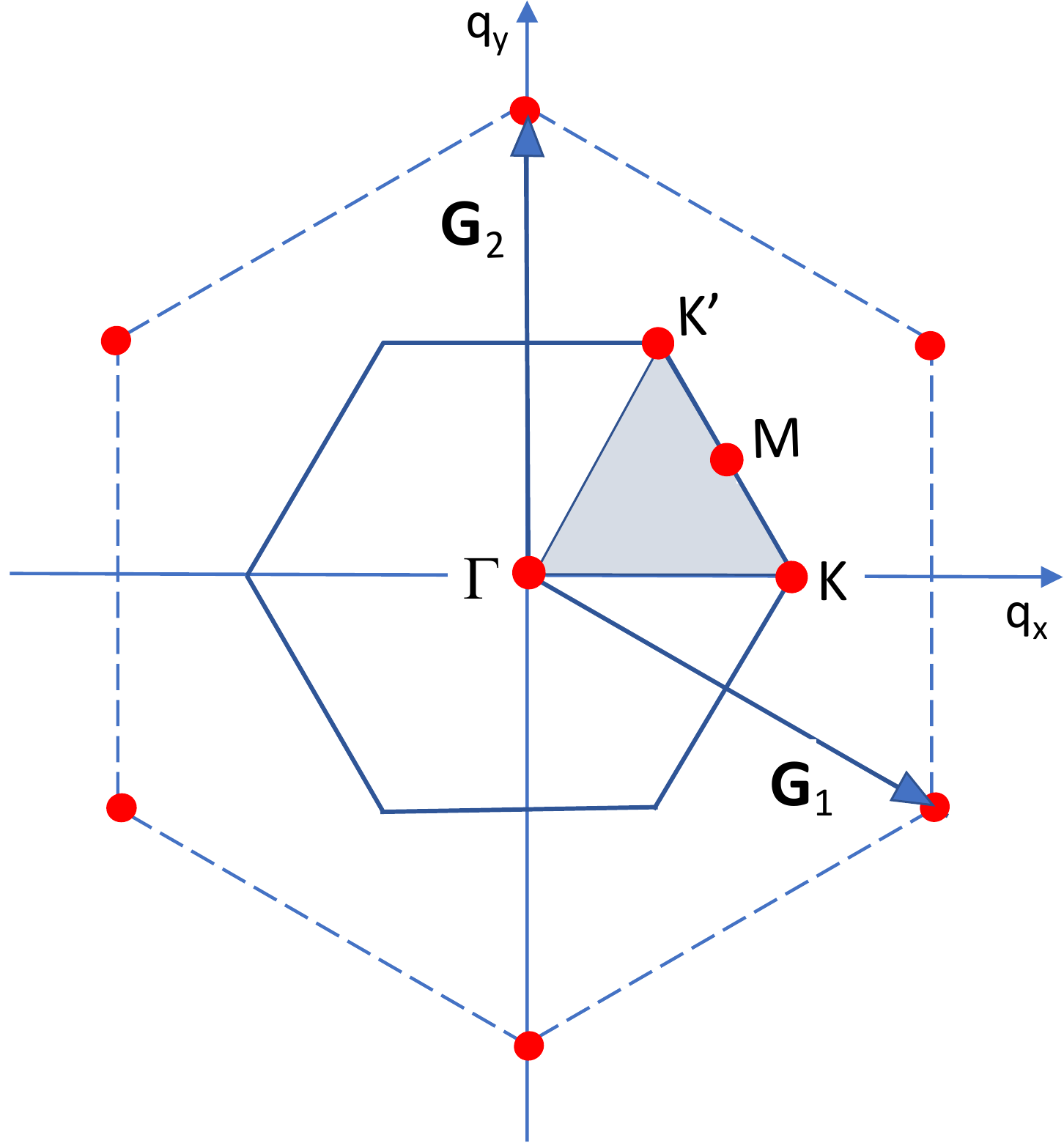}
\caption{The first Brillouin zone (enclosed by solid lines) for the 2D honeycomb lattice.}\label{bzhoney}
\end{figure}
Schematically, $W^c=vP^0v + vP^0vP^0v + ...$ contains the screening from successive particle-hole excitations, each described by the mean-field polarization function, which in position representation reads 
\begin{align}
   &  P^0({\bf rr}';\omega)  = 2 \sum_{{\bf k}n}\sum_{{\bf k}'n'} \psi_{n{\bf k}}^*({\bf r}) \psi_{n'{\bf k}'}({\bf r}) \psi_{n'{\bf k}'}^*({\bf r}') \psi_{n{\bf k}}({\bf r}') \nonumber \\ & \times  \bigg[ \frac{n_{n{\bf k}}(1-n_{n'{\bf k}'})}{\omega \!-\! E_{n'{\bf k}'} + E_{n{\bf k}} + i\delta } - \frac{n_{n'{\bf k}'}(1-n_{n{\bf k}})}{\omega \!-\! E_{n'{\bf k}'} + E_{n{\bf k}} - i \delta } \bigg] .
\end{align}
Here, $n_{n{\bf k}}$ is the occupation of the state $\psi_{n{\bf k}}$ and the factor of $2$ comes from summing over both spin channels. In this work, we perform a one-shot $GW$ calculation. We therefore add a term $[\Delta \mu] \delta_{nn'}$ to Eq. \eqref{Dyson} corresponding to a shift of the chemical potential, to ensure particle number conservation. 
%where the correlated (screening hole) term of the screened Coulomb interaction 
%\begin{align}
%W^c_\text{HF}({\bf q};\omega) & = v({\bf q}) \chi ({\bf q};\omega) v({\bf q}) 
%\end{align}
%depends on the response function $\chi$ within the random-phase approximation,
%\begin{align}
%\chi ({\bf q};\omega) & = P({\bf q};\omega) + P({\bf q};\omega) v({\bf q}) \chi ({\bf q};\omega) .
%\end{align}
%Note that $v({\bf q})$ is calculated within the local approximation. The polarization function is given by 
%(c) $GW$ approximation based on the non-interacting Green function, $g_0({\bf q};\omega) = (\omega - H_0({\bf q}))^{-1}$. Here, $H_0$ is defined from 
%\begin{align*}
%    \Hat{H}^\sigma_0 = \int d{\bf r} \psi_\sigma^\dag ({\bf r}) H_0({\bf r}) \psi_\sigma({\bf r}),
%\end{align*}
%and is recovered from $H$ when removing the mean field of $\hat{V}_{e-e}^\sigma$. The approximation reads 
%\begin{align*}
%G({\bf q};\omega) & = \big( g_0^{-1}({\bf q};\omega) - V_H({\bf q}) - \Sigma^c({\bf q};\omega) + \Delta \mu \big)^{-1}, \\
%\Sigma^c({\bf q};\omega) & \approx i \! \int \frac{d{\bf q}'d\omega'}{(2\pi)^4} g_\text{0}({\bf q}\!+\!{\bf q}';\omega \! + \! \omega') W^c_\text{0}({\bf q}';\omega').
%\end{align*}
%$W^c_0$ is obtained in the same manner as $W^c$ by replacing $g$ by $g_0$ in the expression for $P$. $V_H$ is calculated from the non-interacting density. One main difference between approximation (b) and (c) is thus that the Hartree potential is calculated self-consistently in the former but not in the latter. 
%(Further details are given in Ref. \cite{aryasetiawan2017}\tor{???}) 

\subsection{Computational Details}

The most straightforward way to calculate the derivative of the Bloch functions which enters Eqs. \eqref{mlq}-\eqref{miq}, is to use the perturbation formula\cite{ceresoli2006}
\begin{equation}
\left\vert \frac{\partial u_{n\mathbf{q}}}{\partial q_{i}}\right\rangle
=\sum_{m\neq n}\frac{\big| u_{m\mathbf{q}}\big\rangle \big\langle
u_{m\mathbf{q}}\big| \frac{\partial H({\bf q})}{\partial q_i}\big| u_{n\mathbf{q}%
}\big\rangle }{E_{n\mathbf{q}}-E_{m\mathbf{q}}}, \label{derpert}
\end{equation}
equally valid for insulators as well as metals. This formula implies that $\big\langle \frac{\partial u_{n\mathbf{q}}}{ \partial q_i}\big|u_{n^{\prime }\mathbf{q}} \big\rangle =0$ if $n^{\prime}=n$. For real systems the sum over \textit{all} bands can be tedious, however in tight-binding calculations for model systems the number of states are usually small.

For a topological insulator ($C \neq 0$) it is not possible to define a smooth and continuous gauge in the {\it whole} first BZ,\cite{soluyanov2012} but since $C$ is an observable it must be gauge invariant. The discontinuity can therefore be moved to the BZ boundary by a suitably chosen gauge. We have adopted a random gauge provided by the diagonalization routine used, where $u_{n\mathbf{q}}$ and  $\frac{\partial u_{n\mathbf{q}}}{ \partial q_i}$ have the same phase.

For the practical computation of $G^+_{nn'}({\bf q})$ and its derivatives with respect to $q_x$ and $q_y$ we use the Matsubara formalism at low temperature with a $30 \times 30$ ${\bf q}$ mesh. To converge the computation of $M_\text{orb}$ (Eqs. \eqref{mlq}-\eqref{miq}) we interpolate the Green's function and its derivatives to a non-uniform mesh which is particularly dense at the ${\bf K}$ and ${\bf K}'$ points (see Fig. \ref{bzhoney}). The results do not depend on whether the interpolation is linear or cubic. 

From the static inverse Green's function it is straightforward to obtain the ${\bf q}$-resolved quasi-particle gap, $E_g({\bf q})$, between the occupied and unoccupied bands at wavevector ${\bf q}$ for the correlated system. In particular, we will present, together with $M_\text{orb}$, the Brillouin zone minimum and average gap, $E_g^\text{min}$ and $\langle E_g \rangle$. The computation of $C$ requires a dense ${\bf q}$ mesh in order to capture the singularity. We can still, with our $30\times 30$ mesh, distinguish between the normal and topological insulating phases. In fact, the transition occurs when the gap closes, i.e. it can be traced by $E_g^\text{min}$. 

\section{Results}
We will start by discussing the effect of the staggered potential, $\Delta_\text{AB}$, on the orbital magnetization at complex hopping angle $\varphi=\pi/4$. Then, we study the effect of adding interactions, first with $\Delta_\text{AB}=2$ and then with $\Delta_\text{AB}=0$. For generality, we also present the magnetization in the plane of $U$ and $\Delta_\text{AB}$. Finally, the dependence of $M_\text{orb}$ on $\varphi$ is studied briefly. All results are obtained using $t_1=1$ and $t_2=1/3$, the same values used in the seminal paper by Thonhauser {\it et al}.\cite{thonhauser2005}

\subsection{Non-interacting case - Effect of varying $\Delta_\text{AB}$}
\begin{figure}[h!]
\includegraphics[width=1\linewidth]{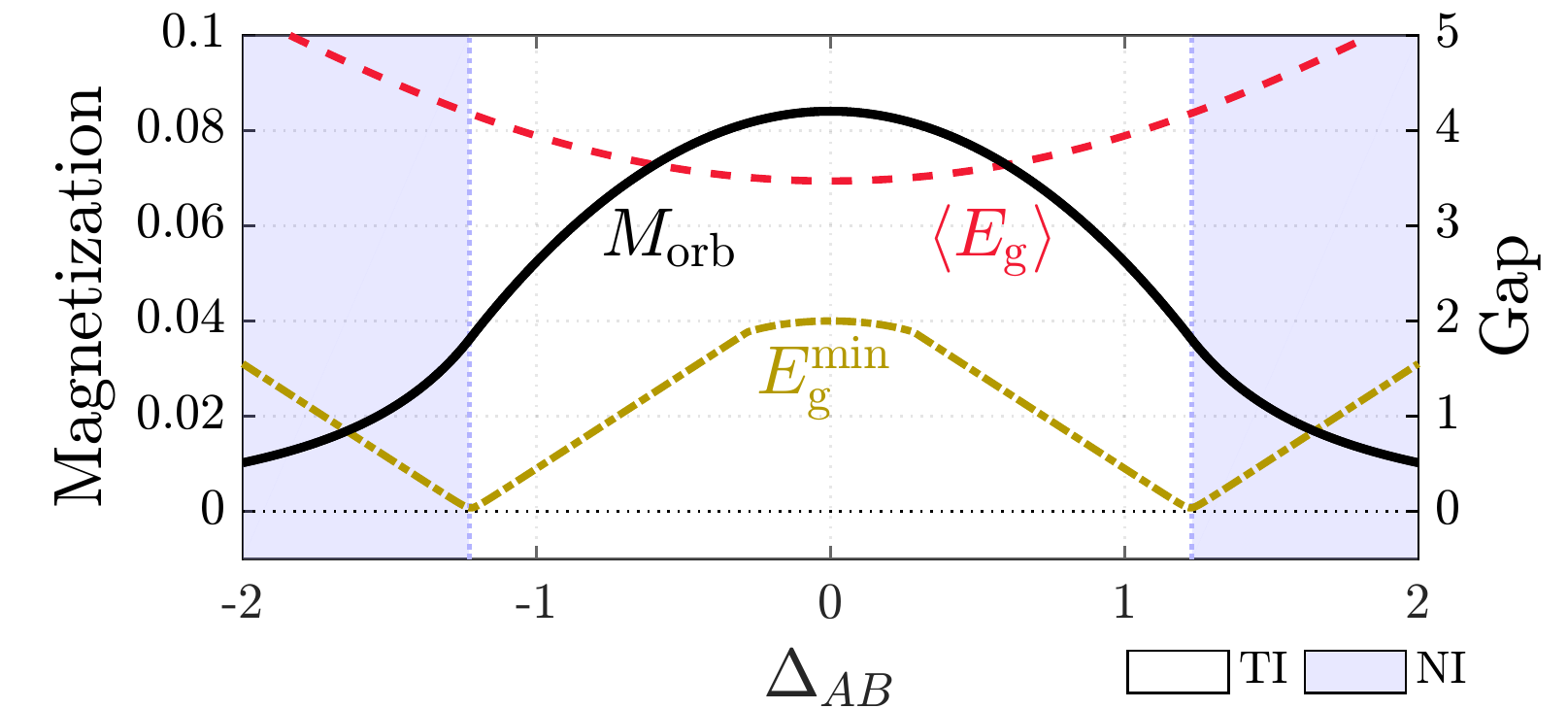}
\caption{Non-interacting $M_\text{orb}$, $E_g^\text{min}$ and $\langle E_g \rangle$ as a function of  $\Delta_\text{AB}$. TI: topological insulator. NI: normal insulator.}\label{plot1}
\end{figure}

The $\Delta_\text{AB}$-dependence of non-interacting $M_\text{orb}$ is shown in Fig. \ref{plot1} together with the minimum and Brillouin zone-averaged ${\bf q}$-resolved gaps, $E_g^\text{min}$ and $\langle E_g \rangle$.  The magnetization decreases monotonously and the average gap increases monotonously when increasing the strength of 
\begin{figure}[h!]
\includegraphics[trim={0 1.5cm 0 0},clip,width=1\linewidth]{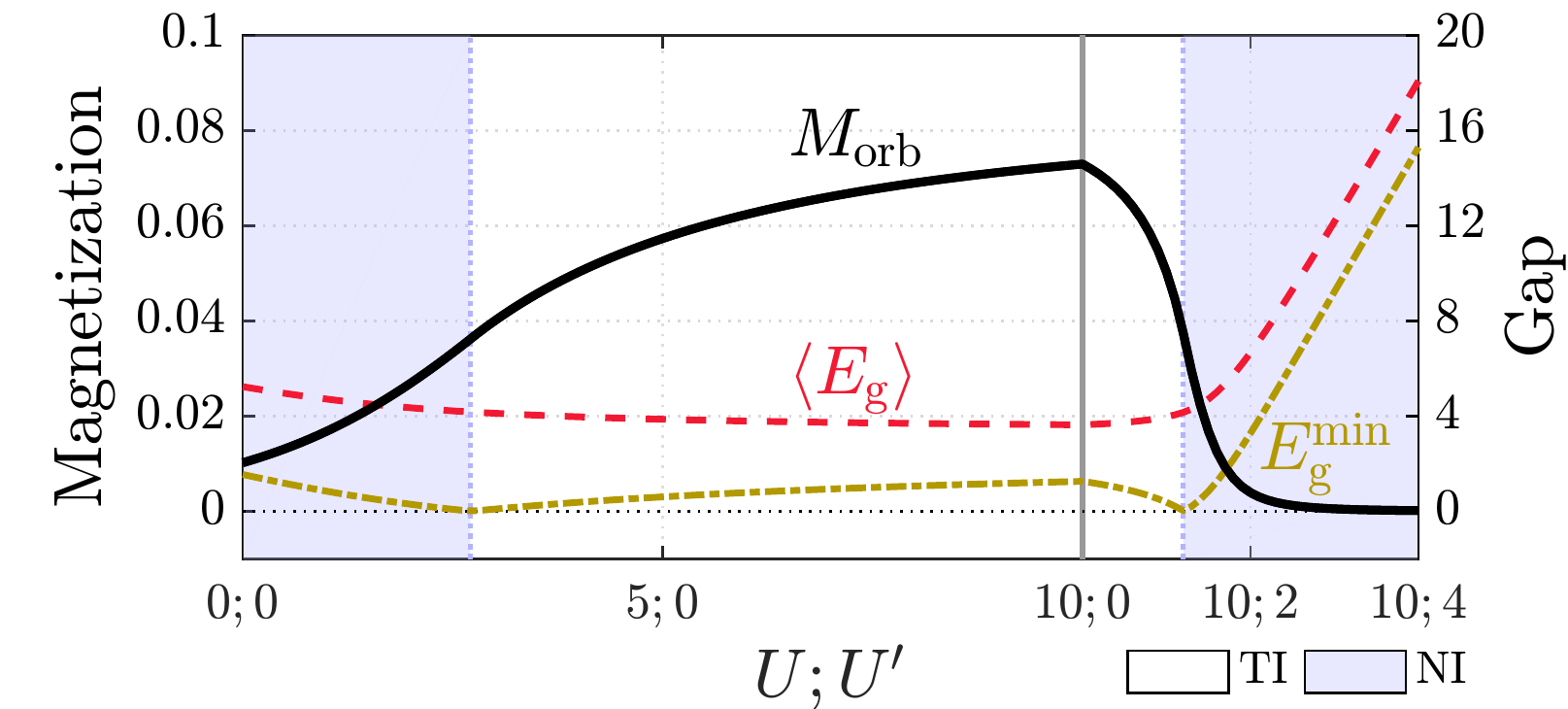}
\includegraphics[width=1\linewidth]{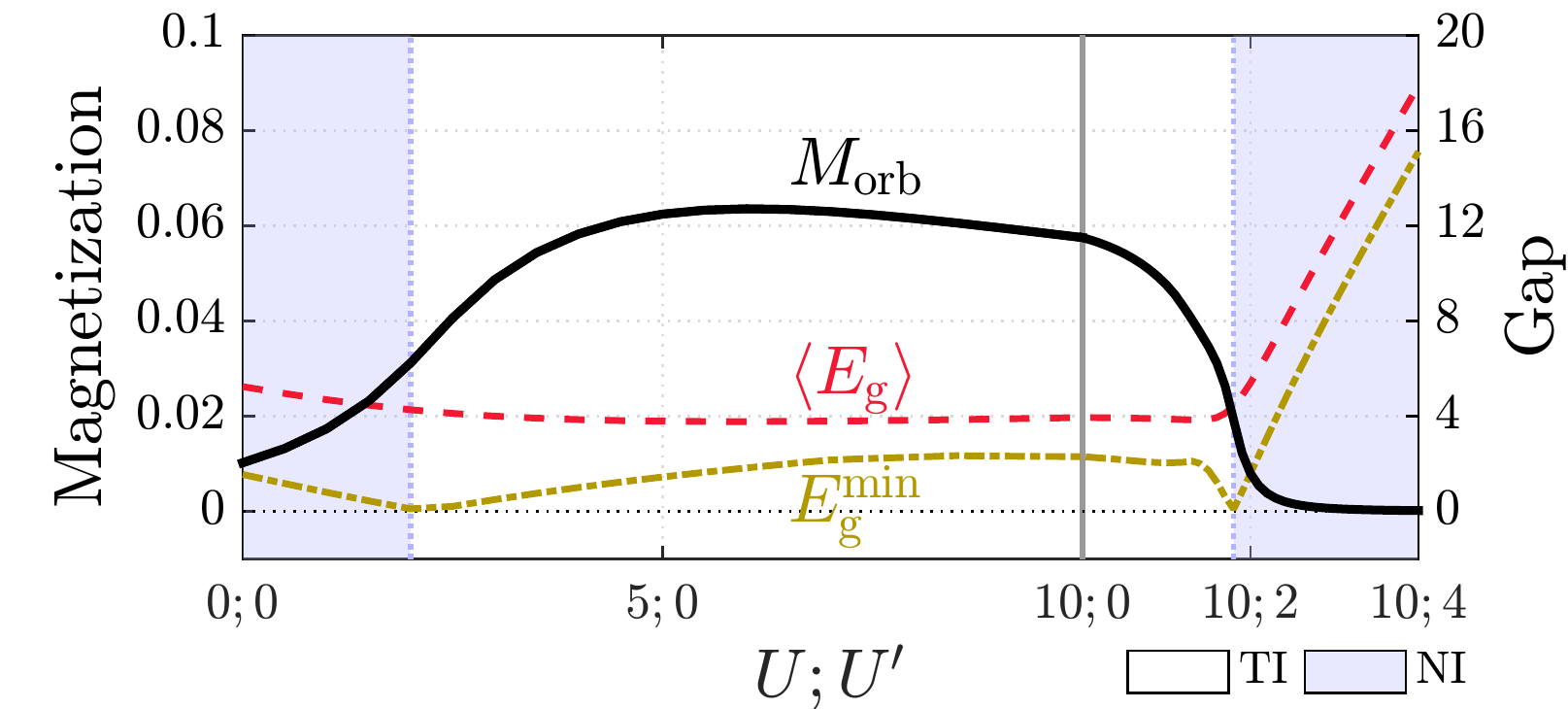}
\caption{$M_\text{orb}$, $E_g^\text{min}$ and $\langle E_g \rangle$ as a function of $U$ and $U'$ with $\Delta_\text{AB}=2$. Top: Hartree-Fock. Bottom: $GW$. }\label{plot2}
\end{figure}
the staggered potential, $|\Delta_\text{AB}|$. This can be understood from the expression 
\begin{align}
M_\text{orb} = & -i \sum_{jk,n'\neq n}  \epsilon_{zjk} \int_{BZ} \frac{d{\bf q}}{(2\pi)^2} \nonumber \; (\varepsilon_{n{\bf q}} + \varepsilon_{n'{\bf q}})\\
& \times \frac{\big\langle u_{n{\bf q}} \big| \frac{\partial H({\bf q})}{\partial q_j}\big| u_{n'{\bf q}}\big\rangle \big\langle u_{n'{\bf q}} \big| \frac{\partial H({\bf q})}{\partial q_k} \big| u_{n{\bf q}}\big\rangle}{(\varepsilon_{n{\bf q}} - \varepsilon_{n'{\bf q}})^2} , \label{Malt}
\end{align}
which holds for non-interacting insulators, for which $G_{nn^{\prime}}(\mathbf{q}%
)=i\delta_{n^{\prime}n}f(E_{n\mathbf{q}})$ and $\partial G_{nn^{\prime}%
}(\mathbf{q})/\partial q_{i}=0$. Note that band $n$ is occupied and $n'$ unoccupied. The one-particle energy is measured relative to $\mu$ i.e $\varepsilon_{n{\bf q}} = E_{n{\bf q}} - \mu$.
If we imagine that the gap $E_g({\bf q})=E_{n'{\bf q}}-E_{n{\bf q}}$ is independent of ${\bf q}$ and thus given by $\langle E_g\rangle$, then a small $\langle E_g \rangle$ implies a large $M_\text{orb}$ and vice versa. This simplified picture is useful if the inverse of $E_g({\bf q})$ is only weakly ${\bf q}$ dependent, which holds when we are far from phase transitions. The non-interacting value of $M_\text{orb}$ at $\Delta_\text{AB}=2$ matches the value of approximately $0.01$ in an earlier work by Thonhauser {\it et al}.\cite{thonhauser2005} 

For small $|\Delta_\text{AB}|$, the system starts out as a topological insulator but at $|\Delta_\text{AB}| \approx 1.2$ a topological phase transition occurs and the system becomes a normal insulator, with $C=0$. This transition is accompanied by the closing and reopening of the gap, $E_g^\text{min}$. 

\subsection{Effect of correlations with a non-zero $\Delta_\text{AB}$}
\begin{figure}[h!]
\includegraphics[width=0.98\linewidth]{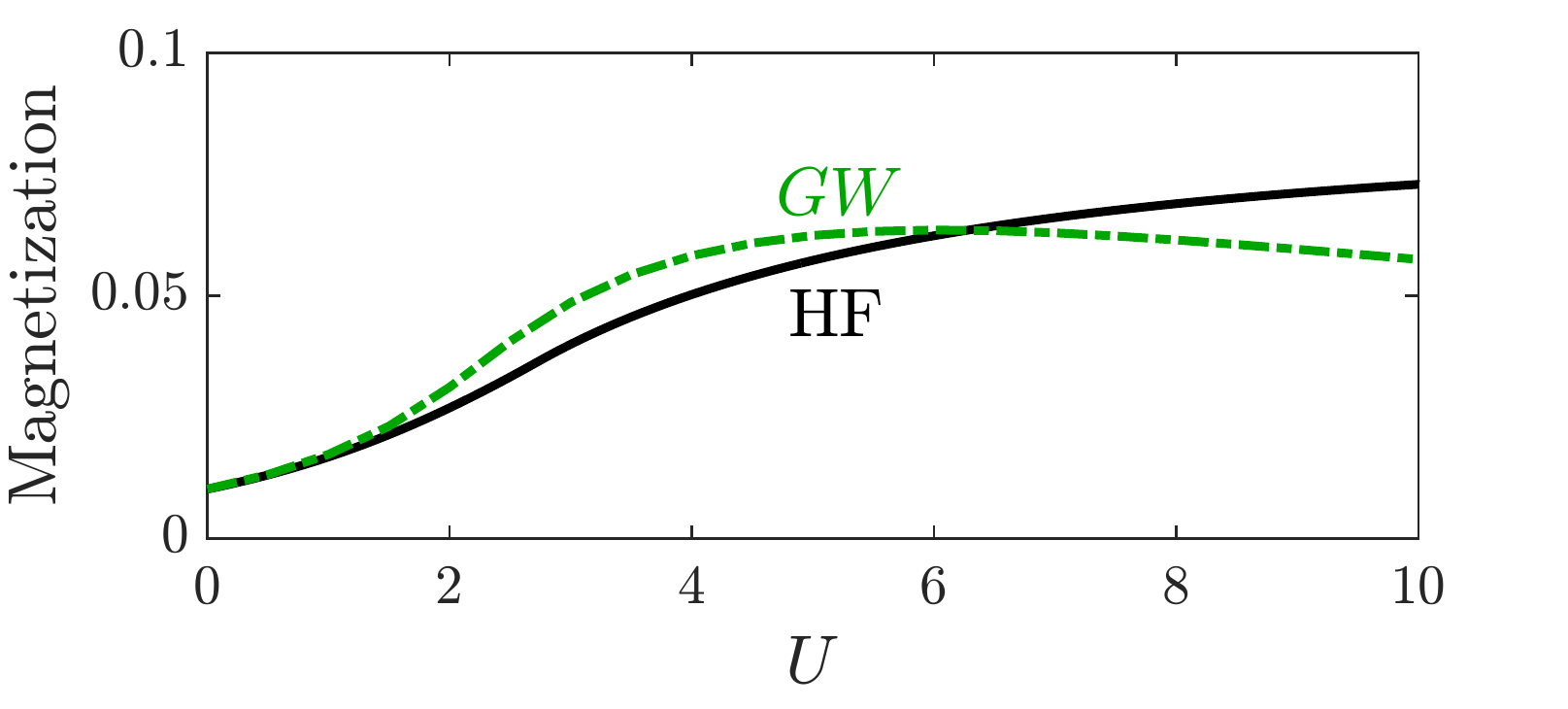}
\caption{Closer look on $M_\text{orb}$ as a function of $U$ within the Hartree-Fock and $GW$ approximation. $\Delta_\text{AB}=2$. }\label{plotcompare}
\end{figure}

We now fix the staggered potential to $\Delta_\text{AB}=2t_1=2$. $M_\text{orb}$ is plotted in Fig. \ref{plot2} versus the on-site and nearest-neighbor direct Coulomb integrals, $U$ and $U'$, within the Hartree-Fock and $GW$ approximation. As a complement to Fig. \ref{plot2}, we present $M_\text{orb}$ in the two approximations in an overlay graph in Fig. \ref{plotcompare}. 
%\begin{figure}[h!]
% \includegraphics[width=0.98\linewidth]{atry.eps}
%\caption{COMMENT: Looks weird in the colorbar (white diagonal line) because of ''updates'' (downdates) of matlab since 2014. %We have to choose between high quality vector graphics or low quality without the diagonal white line.  }
%\end{figure}
\begin{figure}[h!]
\includegraphics[width=0.98\linewidth]{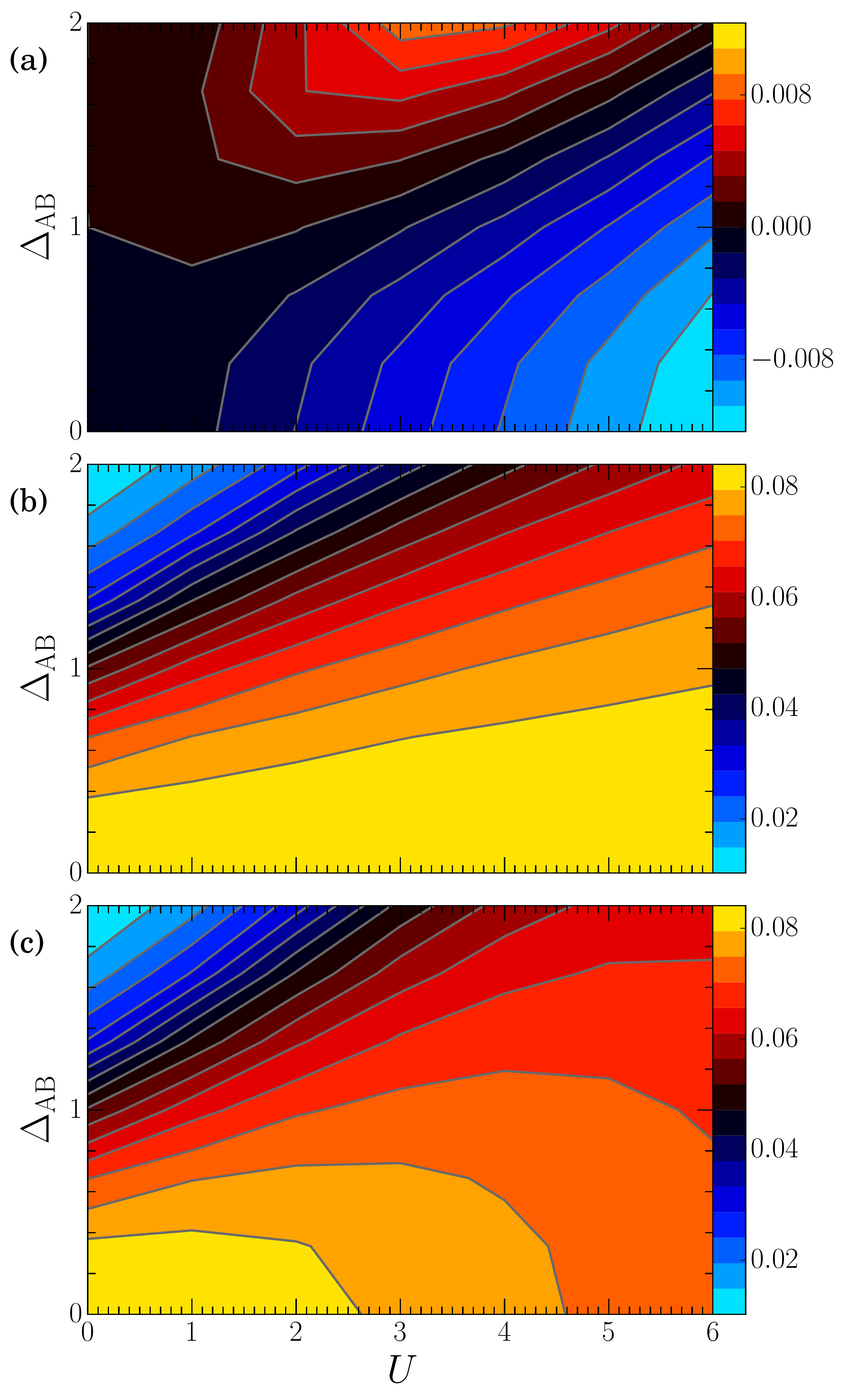}
\caption{The effect of correlations. (a) Difference between the $GW$ and the Hartree-Fock magnetization. The correlation boost is clearly seen for large $\Delta_\text{AB}$ and intermediate $U$. (b) Hartree-Fock magnetization. (c) $GW$ magnetization. }\label{plotplane}
\end{figure}
%\begin{figure}[h!]
%\includegraphics[width=1\linewidth]{OHFD2.pdf}
%\caption{One-shot Hartree-Fock calculation. $M_\text{orb}$, $E_g^\text{min}$ and $\langle E_g \rangle$ as a function of $U$ and $U'$ with %$\Delta_\text{AB}=2$.}\label{plot3}
%\end{figure}

The HF results show that the mean-field effect of increasing $U$ to infinity is to recover the non-interacting $M_\text{orb}$ at $\Delta_\text{AB}=0$ in Fig. \ref{plot1}, in other words, $U$ counteracts the staggered potential. The effect of $U'$, which instead enhances it, is a decrease in $M_\text{orb}$.

The $GW$ approximation yields a slightly larger $M_\text{orb}$ than the Hartree-Fock approximation for small $U$. This increase originates from dynamical and non-local charge correlations, encompassed by the polarization function $P$, which makes the $GW$ Green's function non-diagonal in the band index, $n$, when $U>0$. The staggered potential, which modified the band energies $E_{n{\bf q}}$ that enter the polarization function, has a non-trivial effect on the self-energy. However, a slight decrease in the average gap, $\langle E_g \rangle$, can be observed compared to Hartree-Fock, which means that the charge fluctuations reduce the band gap. We have already noticed, in the non-interacting case, that a small band gap is associated with a large value of $M_\text{orb}$, but in the interacting case the entire dynamics has to be considered for a complete understanding, since $G_{nn'}^+({\bf q})$ is obtained by integrating over all frequencies in Eq. \eqref{intfreq}. 

We have confirmed that the boost of the magnetization for small and intermediate $U$ is caused by the last terms in Eq. \eqref{mlq} and \eqref{miq} respectively, which are due to interband correlations ($n'\neq n$), with no non-interacting counterpart. Without these terms the magnetization would rapidly vanish when turning on the correlations. The correlation boost is seen in Fig. \ref{plotplane}, where the difference between the $GW$ and Hartree-Fock magnetization is displayed in the plane of $U$ and $\Delta_\text{AB}$ together with the magnetization within each approximation. We see that the effect of correlations depends strongly on $\Delta_\text{AB}$, and that boost occurs when the staggered potential dominates the nearest-neighbor hopping, i.e. when $\Delta_\text{AB}>1$. 

\subsection{Effects of correlations with $\Delta_\text{AB}=0$}
\begin{figure}[h!]
\includegraphics[trim={0 1.5cm 0 0},clip,width=1\linewidth]{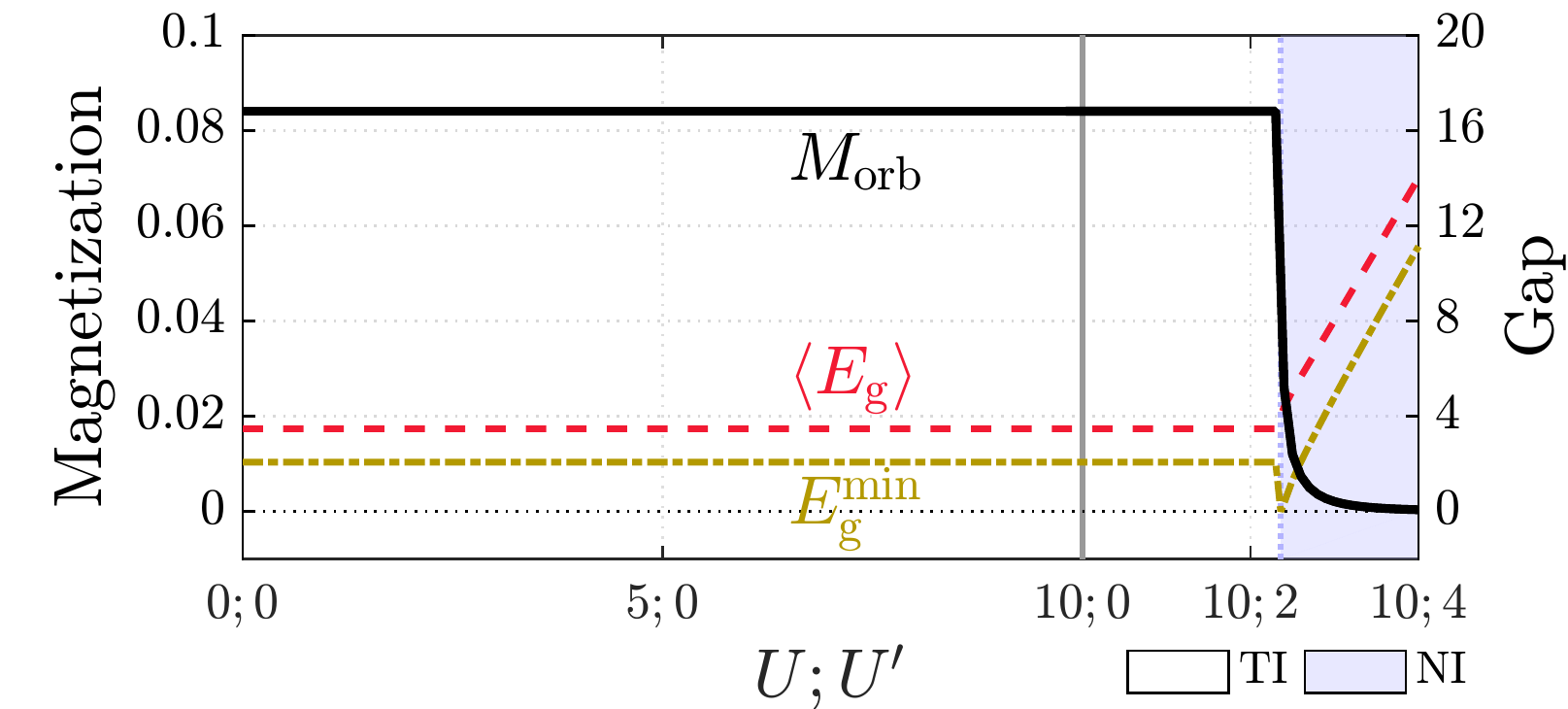}
\includegraphics[width=1\linewidth]{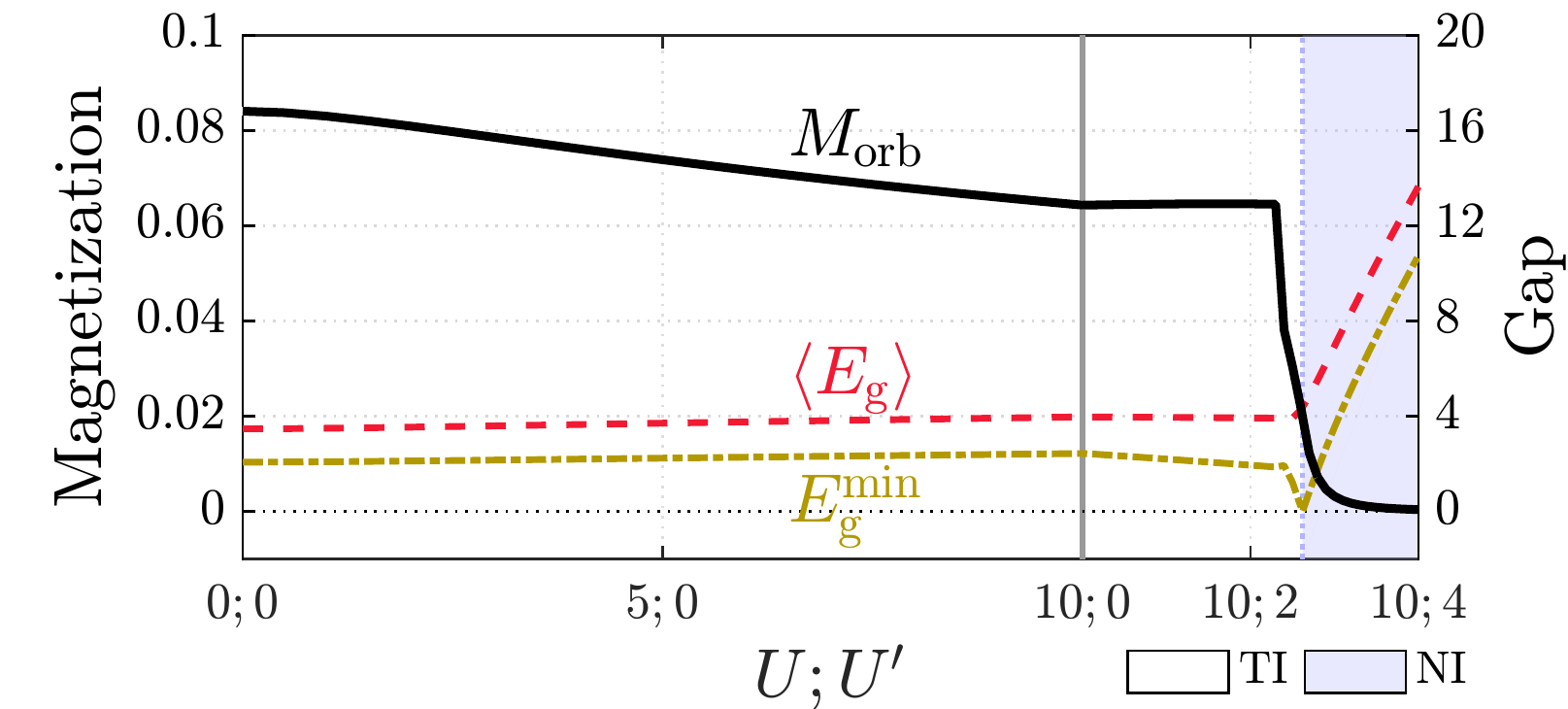}
\caption{$M_\text{orb}$, $E_g^\text{min}$ and $\langle E_g \rangle$ as a function of $U$ and $U'$ with $\Delta_\text{AB}=0$. Top: Hartree-Fock. Bottom: $GW$. }\label{plot4} 
\end{figure}

\begin{figure}[h!]
\includegraphics[width=1\linewidth,right]{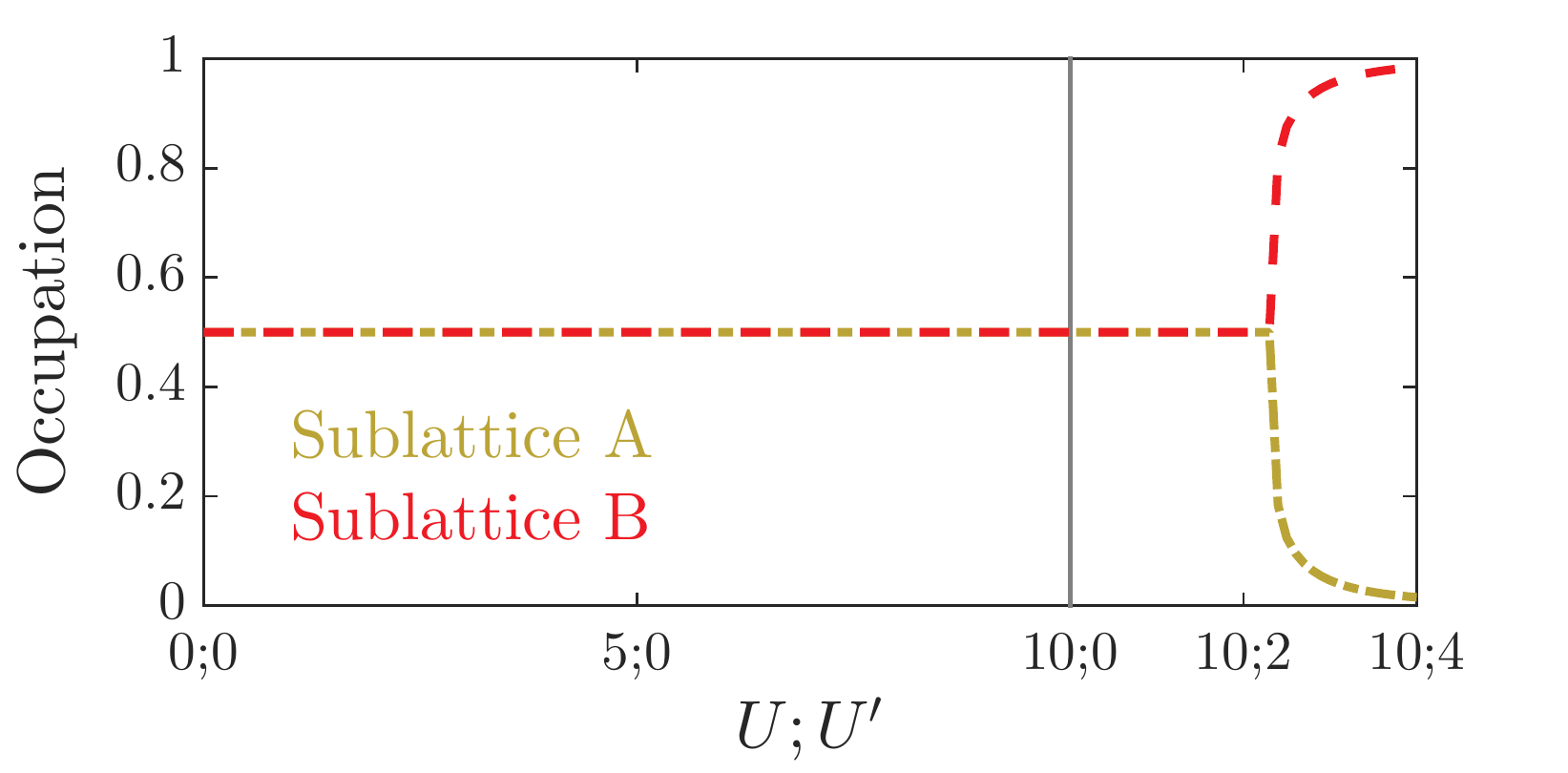}
\caption{Occupation of sublattice A and B as a function of $U$ and $U'$ within the Hartree-Fock approximation, with $\Delta_\text{AB}=0$. Charge segregation occurs when $U'$ exceeds a critical value. }\label{plot42} 
\end{figure}

Results analogous to those in Fig. \ref{plot2} are presented in Fig. \ref{plot4}, with $\Delta_\text{AB}=0$. Within the Hartree-Fock approximation, the magnetization is independent of $U$. This is because the staggered potential is absent, so the on-site repulsion yields the same constant energy shift to both bands. When increasing $U'$ a sharp phase transition takes place after which $M_\text{orb}$ decays rapidly. The reason for this sudden decrease of $M_\text{orb}$ is seen in Fig. \ref{plot42}, where the occupation of sublattice A and B is plotted versus $U$ and $U'$ within the Hartree-Fock approximation, for $\Delta_\text{AB}=0$. For a fixed value of $U=10$, charge segregation occurs for a sufficiently large value of $U'$, after which all electrons very quickly end up occupying only sublattice B. This introduces a purely electronic counterpart of the ionic potential, $\Delta_\text{AB}$, and naturally, the effect is a reduced $M_\text{orb}$.

Returning to Fig. \ref{plot4}, we see that in the $GW$ approximation
 $M_\text{orb}$ reduces with $U$, as opposed to the case $\Delta_\text{AB}=2$ in Fig. \ref{plot2}. With $\Delta_\text{AB}=0$, the mean-field potential vanishes (except for large values of $U'$) and does therefore not affect the charge fluctuations. As mentioned before, the behavior of $M_\text{orb}$ can not be completely understood in terms of the gap, but it is worth stressing that $\langle E_g \rangle$ indeed increases with $U$. 
\begin{figure}[h!]
\includegraphics[width=1\linewidth]{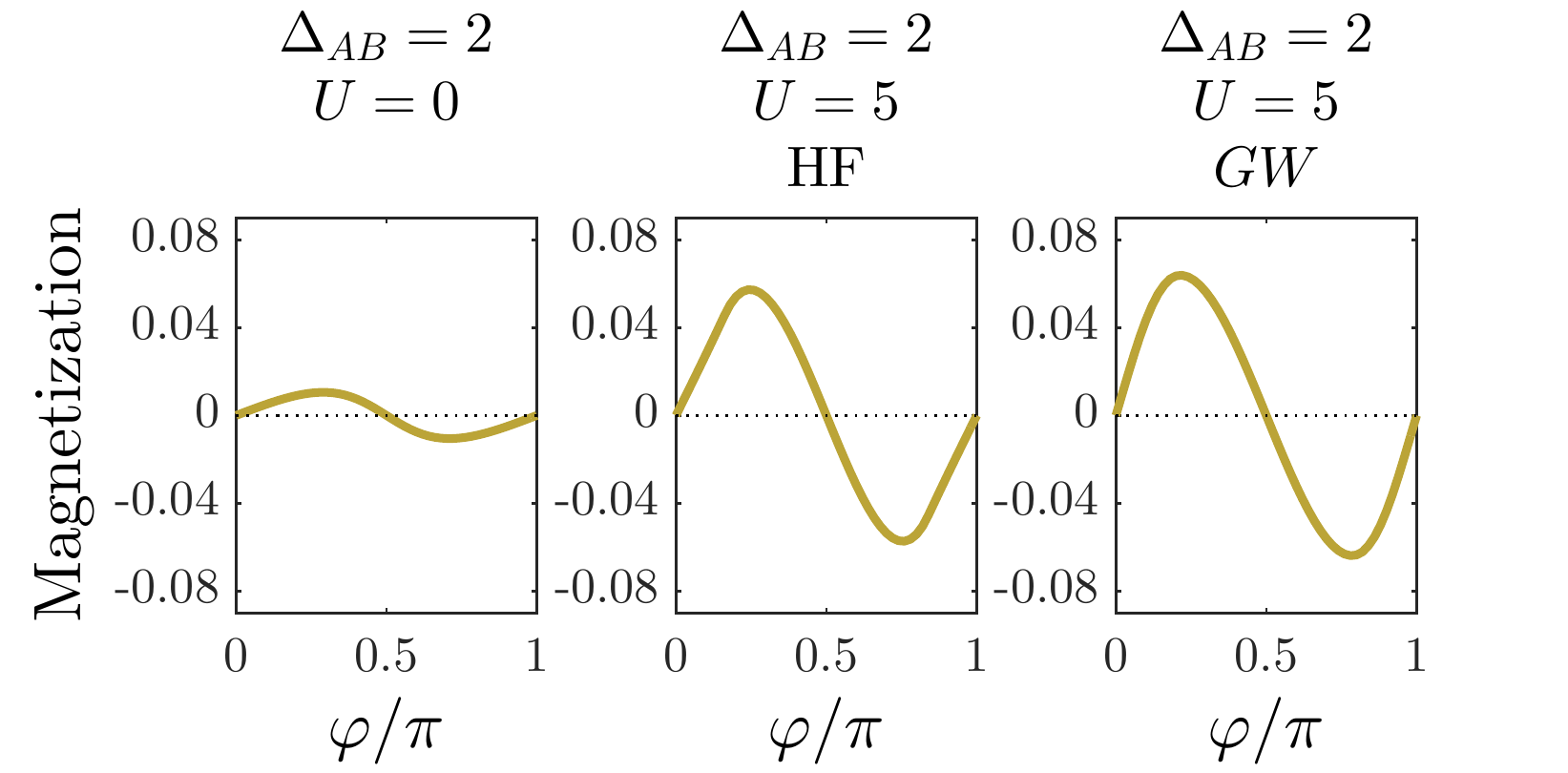}
\caption{$M_\text{orb}$ as a function of the complex hopping phase $\varphi$. $U'$ is set to zero.}\label{plot5} 
\end{figure} 
\subsection{Effect of varying $\varphi$}
The dependence of $M_\text{orb}$ on the complex hopping phase $\varphi$ is presented in Fig. \ref{plot5} for $\Delta_\text{AB}=2$ in the non-interacting limit and for $U=5$, $U'=0$ within the Hartree-Fock and the $GW$ approximations. The curve for $U=0$ is identical to Fig. 2 in Ref. \citen{thonhauser2005} and displays a characteristic sinus-like $\varphi$ dependence.\cite{thonhauser2005,thonhauser2006} This behavior survives in the interacting case in both approximations. Since the maximum value is obtained approximately at $\varphi=\pi/4$, independently of the interaction strength and the approximation, the $M_\text{orb}$ presented in Fig. \ref{plot1}-\ref{plot4} can be interpreted as the maximum possible magnetization for fixed $\Delta_\text{AB}$, $U$ and $U'$. 

\section{Summary and Conclusions}
The effect of charge correlations on the orbital magnetization has been studied in the 2D Haldane-Hubbard model by adding the $GW$ self-energy. To the best of our knowledge, this is the first study on orbital magnetization where the screening has been treated microscopically. Qualitatively, we are lead to distinguish between two quite different situations.

(i) When the staggered potential dominates the nearest-neighbor hopping we find that the main effect of $GW$ correlations, for small values of the the on-site $U$ and vanishing $U'$, is an increase of the orbital magnetization compared to the Hartree-Fock approximation. In fact, the non-interacting modern theory of orbital magnetization is known to underestimate $M_\text{orb}$ in materials outside the realm of the Haldane-Hubbard model, such as ferromagnetic transition metals. The explanation is hidden in the frequency-integral of Eq. \eqref{intfreq}, which depends on the entire dynamics, in particular, both the renormalization of the two bands, as well as the gap. In the non-interacting limit, the correlation boost of $M_\text{orb}$ can be understood mainly from the narrowing of the quasiparticle gap, using Eq. \eqref{Malt}.\cite{ceresoli2010} 

(ii) When the nearest-neighbor hopping dominates the staggered potential, the $GW$ correlations are instead found to yield a decrease in the magnetization for small values of $U$ and vanishing $U'$. For $\Delta_\text{AB}=0$, inversion symmetry is recovered and the polarization function used to calculate the Green's function within the one-shot $GW$ approximation becomes independent of $U$. For large enough values of $U'$, charge segregation sets in, resulting in a purely electronic equivalent of the staggered potential which breaks inversion symmetry, resulting in a drop in $M_\text{orb}$ both within the Hartree-Fock and the $GW$ approximation.

\begin{acknowledgments}
This work was supported by the Swedish Research
Council VR (T. J. Sj\"{o}strand and F. Aryasetiawan). One of us (K. Karlsson) would like to thank J. Kanski for fruitful discussions. 
\end{acknowledgments}

\end{document}